\documentclass[12pt,twoside]{article}

\usepackage{cmp2e}

\title[Symmetries in the Physics of SCES]{Symmetries in the Physics of\\
Strongly Correlated Electronic Systems\thanks{It is our great pleasure
to dedicate this article to Professor \emph{Ihor Stasyuk} on the
occasion of his $60^{th}$ birthday.}}

\author[F.~Mancini and A.~Avella]{F.~Mancini and A.~Avella}
\address{Universit\`a degli Studi di Salerno -- Unit\`a INFM di Salerno\\
Dipartimento di Scienze Fisiche ``E.R. Caianiello''\\ 84081 Baronissi,
Salerno, Italy}

\begin{document}

\maketitle

\begin{abstract}
Strongly correlated electron systems require the development of new
theoretical schemes in order to describe their unusual and unexpected
properties. The usual perturbation schemes are inadequate and new
concepts must be introduced. In our scheme of calculations, the
Composite Operator Method, is possible to recover, through a
self-consistent calculation, a series of fundamental symmetries by
choosing a suitable Hilbert space.
\pacs 71.10.-w, 71.10.Fd, 71.27.+a.
\end{abstract}

The discovery of new materials with a large variety of unusual and
unexpected properties \cite{RevExp} has opened a new \emph{era} in the
physics of Condensed Matter; new theoretical schemes must be developed
\cite{RevThe}. The most important characteristic of these new systems is
a strong correlation among the electrons that makes inapplicable
classical schemes based on the band picture. It is necessary to pass
from a \emph{single-electron} physics to a \emph{many-electron} physics,
where the dominant part will be the correlations among the electrons.
Usual perturbation schemes are inadequate and new concepts must be
introduced.

Let us consider a certain Hamiltonian
\begin{equation}
H=H\left[\varphi_1(x),\dots,\varphi_n(x)\right]
\label{eq1}
\end{equation}
the set $\{\varphi_i\}$ denotes band-electron fields. Due to strong
correlation effects the properties of the original electrons
$\{\varphi_i\}$ are drastically changed; new excitation modes will
appear and determine most of the observed properties of the system. It
is natural to identify a new set of elementary excitations $\{\psi_i\}$
as basis for perturbative schemes. These excitations, constructed from
the original electron set (in this sense we call them composite fields),
are created by the interactions among the electrons; therefore, their
properties will be determined by the dynamics, the symmetries of the
model, the boundary conditions and must be computed in a self consistent
way \cite{COM}. This aspect introduces a new richness into the theory
and will allow us to realize the dynamics in the proper Hilbert space
where the physical symmetries are conserved. On the other hand, we know
from experiments that highly correlated systems exhibit an incredible
variety of behaviors. It would be very hard to describe such a
complexity using the original fields, unless the exact solution of the
model is available. The presence of new excitations, composite fields,
introduces into the theory the possibility to accommodate the
multifariousness of experimental properties.

A theory built on the basis of new excitation modes is, by construction,
a self-consistent theory, and a procedure must be fixed. In particular,
we must answer to the following list of questions:
\begin{enumerate}
\item the identification of the fundamental set;
\item the statistics and the properties of the new fields;
\item the symmetry and the dynamics in terms of the new fields;
\item the representation where the new fields are realized.
\end{enumerate}

We will now try to formulate a scheme of calculations in which an answer
to the previous questions can be found. Then, by considering a
particular model, we will present a practical realization of the
theoretical scheme.

The new fields $\{\psi_i\}$ are generated by the interactions among the
bare fields; then, it is naturally to choose the new set as the one
which naturally appears through the equations of motion. The evolution
of the original fields is described by the Heisenberg equation
\begin{equation}
i\frac{\partial}{\partial t}\varphi_i(x)=\left[\varphi_i(x),H\right]
=J_i\left[\varphi(x)\right]
\label{eq2}
\end{equation}

Such an equation generates new fields $\{J_i[\varphi(x)]\}$, constructed
as combinations of the bare fields. By starting from these fields and by
considering the new Heisenberg equations
\begin{equation}
i\frac{\partial}{\partial t}\psi_i(x)=\left[\psi_i(x),H\right]
\label{eq3}
\end{equation}
we generate an infinite hierarchy of composite fields. It is naturally
impossible to solve the infinite system of equations and some truncation
procedure must be adopted. Let us consider a $n$-component field
\begin{equation}
\psi=\left(
 \begin{array}{c}
  \psi_1 \\ \vdots \\ \psi_n
 \end{array}
\right)
\label{eq4}
\end{equation}
and let us choose the first $n-1$ fields such as
\begin{equation}
i\frac{\partial}{\partial t}\psi_i(x)=\left[\psi_i(x),H\right]
=\sum_{j=1}^{i+1}\gamma_{ij}(-i\nabla)\,\psi_j(x)
\hspace{1cm} 1 \le j \le n-1
\label{eq5}
\end{equation}

The $n^{th}$-field $\psi_n(x)$ is determined by the field equation of
$\psi_{n-1}(x)$. The matrix $\gamma(-i\nabla)$ is completely determined
by the dynamics. Then, we linearize the Heisenberg equation by writing
\begin{equation}
i\frac{\partial}{\partial t}\psi(x)=\varepsilon(-i\nabla)\,\psi(x)
\label{eq6}
\end{equation}
where the eigenvalue or energy matrix $\varepsilon$ is self-consistently
calculated by means of the equation
\begin{equation}
\varepsilon(-i\nabla_x)\left\langle\left\{\psi(\mathbf{x},t),
\psi^\dagger(\mathbf{y},t)\right\}\right\rangle
=\left\langle\left\{\left[\psi(\mathbf{x},t),H\right],
\psi^\dagger(\mathbf{y},t)\right\}\right\rangle
\label{eq7}
\end{equation}

The symbol $\langle\cdots\rangle$ denotes the thermal average.
Derivative operators as $\lambda(-i\nabla)$ are defined through the
relation
\begin{equation}
\lambda(-i\nabla)\,f(\mathbf{x})
=\int d^dy\,\lambda(\mathbf{x},\mathbf{y})\,f(\mathbf{y})
\label{eq8}
\end{equation}

The rank of the energy matrix is equal to $n$, the number of components
of the vector $\psi(x)$. When there is translational invariance we can
invert Eq.~(\ref{eq7}) and it is easy to see that
\begin{eqnarray}
\varepsilon_{ij}=\left\langle\left\{\left[\psi_i(x),H\right],
\psi_l^\dagger(y)\right\}\right\rangle
\left\langle\left\{\psi_l(x),
\psi_j^\dagger(y)\right\}\right\rangle^{-1}
=\gamma_{ij} \nonumber\\
1 \le i \le n-1, 1 \le j \le n
\label{eq9}
\end{eqnarray}

This approximation corresponds to the $n$-pole expansion of the Green's
function where finite life-time contributions are neglected. It has been
proved \cite{Spectral} that in this approximation the choice (\ref{eq5})
for the composite operators leads to the conservation of the spectral
moments. In particular, the first $2(n-i+1)$ spectral moments for the
field $\psi_i [1 \le i \le n-1]$ are conserved. This is an important
property when we recall that the spectral moments are related to the
spectral density function of the
\emph{single-particle} propagators. Also, as shown in
Ref.~\cite{Spectral}, the choice (\ref{eq5}) leads to an equivalence
between the $n$-pole approximation and the spectral density approach
\cite{Poles}, although very different results are obtained when
different procedures for the self-consistency are used \cite{Two}.

In general the composite fields will not satisfy canonical
anticommutation relations and their algebra must be calculated starting
from the canonical algebra of the electron fields. Owing to this fact,
the Wick theorem and the standard perturbation schemes cannot be
applied. Examples of the new algebra will be presented in the second
part of this article.

The properties of the new fields are fixed by a series of parameters
which must be self-consistently calculated. These parameters are
expressed as expectation values of composite fields. When the composite
fields belong to the set they can be expressed in terms of the
\emph{single-particle} Green's function and calculated by a series of
coupled self-consistent equations.

However, it may happens that some of the parameters are expressed as
expectation values of higher-order composite fields that do not belong
to the basic set. In this case, owing to the approximation considered,
the parameters are not strictly bound by the dynamics and there is a
freedom in the procedure to fix them. At this level the powerfulness of
the scheme manifests itself: one can use this freedom to choose the
right representation. In the construction of a physical theory we must
distinguish two levels. On one side we have the microscopic level where
we are concerned with particles. The basic ingredients are the
Heisenberg fields which together with the canonical commutation
relations describe the dynamics. The physical laws (the equations of
motion, the conservation laws, the symmetry principles) are expressed as
relations among the operators. On the other side we have the macroscopic
world where we are concerned with average values of operators. At the
level of observation the physical laws manifest themselves as relations
among matrix elements, and a suitable choice of the Hilbert space must
be made. When some approximation is introduced the states are not the
exact eigenstates of the Hamiltonian; the expectation values are also
not the exact ones. As a consequence, the relations among the operators
are generally not conserved when the expectation values are calculated.
A striking example of this is the violation of the Pauli principle. A
convenient way to take care of it is to operate in the representation of
second quantization where the Pauli principle manifests through the
algebra. It is known \cite{Rowe} that in most of the approximation
schemes this symmetry is violated when matrix elements are considered.
Other examples of symmetries will be considered later.

The point of view adopted in this approach is that we can use the
freedom in the procedure to fix the self-consistent parameters in such a
way to recover the symmetries violated by the approximation. In general,
a model exhibits many different symmetries and there will be a relation
between the number of composite fields and the number of symmetries that
can be recovered. On the physical ground one must choose which
symmetries are the most important to be satisfied. In a physics
dominated by strong electron correlations the Pauli principle plays a
crucial role, and it is extremely relevant that the related symmetries
be treated in a correct way. Therefore, in our scheme the attention is
firstly put to the Pauli principle; once this is accommodated, the
attention is devoted to other symmetries.

As an illustration of the scheme we shall now consider the Hubbard model
\cite{Hub}. In a standard notation this model is described by the
following Hamiltonian
\begin{equation}
H=\sum_{ij}\left(t_{ij}-\mu\,\delta_{ij}\right)c^\dagger(i)\,c(i)+U\sum_i
n_\uparrow(i)\,n_\downarrow(i)
\label{eq10}
\end{equation}
$c(i),\,c^\dagger(i)$ are annihilation and creation operators for
electrons at site $i$ in the spinor notation
\begin{equation}
c(i)=\left(
\begin{array}{c}
c_\uparrow(i)\\c_\downarrow(i)
\end{array}
\right)
\hspace{1cm}
c^\dagger(i)=\left(c^\dagger_\uparrow(i),c^\dagger_\downarrow(i)\right)
\label{eq11}
\end{equation}
$n_\sigma=c_\sigma^\dagger(i)\,c_\sigma(i)$ is the number operator of
electrons with spin $\sigma=\left(\uparrow,\downarrow\right)$ at the
$i^{th}$ site. $\mu$ is the chemical potential and is introduced in
order to control the band filling $n$. For a two-dimensional squared
lattice and by restricting the analysis to first nearest neighbors, the
hopping matrix $t_{ij}$ has the form
\begin{eqnarray}
t_{ij}&=&-4t\,\alpha_{ij}=-4t\frac1N\sum_\mathbf{k}
e^{i\,\mathbf{k}\cdot\left(\mathbf{R}_i-\mathbf{R}_j\right)}\,\alpha(\mathbf{k})
\label{eq12}
\\
\alpha(\mathbf{k})&=&\frac12\left[\cos(k_xa)+\cos(k_ya)\right]
\label{eq13}
\end{eqnarray}
$a$ being the lattice constant. In addition to the band term, the model
contains an interaction term which approximates the interaction among
the electrons. In the simplest form of the Hubbard model, the
interaction is between electrons of opposite spin on the same lattice
site; the strength of the interaction is described by the parameter $U$.

The electron field $c(i)$ satisfies the Heisenberg equation
\begin{equation}
i\frac{\partial}{\partial t}c(i)=-\mu\,c(i)-4t\,c^\alpha(i)+U\,\eta(i)
\label{eq14}
\end{equation}
where
\begin{equation}
c^\alpha(i)=\sum_j\alpha_{ij}\,c(j)
\label{eq15}
\end{equation}
is the electron field on the nearest neighbor sites. We see that the
dynamics has generated the composite field
\begin{equation}
\eta(i)=c(i)\,n(i)
\label{eq16}
\end{equation}
The Heisenberg equation for this field will generate a new higher order
composite field. The process does not stop and an infinite number of
composite fields will be generated. By following the procedure mentioned
above, we close the hierarchy by considering n fields and we construct
the vector composite field as described by Eqs.~(\ref{eq4}) and
(\ref{eq5}). For the specific case, we consider the three-component
field
\begin{equation}
\psi(i)=\left(
 \begin{array}{c}
  \psi_1(i) \\ \psi_2(i) \\ \psi_3(i)
 \end{array}
\right)
\label{eq17}
\end{equation}
where
\begin{eqnarray}
\psi_1(i)&=&\xi(i)=c(i)\left[1-n(i)\right]\nonumber\\
\psi_2(i)&=&\eta(i)=c(i)\,n(i)\\
\psi_3(i)&=&\pi(i)=\frac12\sigma^\mu\,n_\mu(i)\,c^\alpha(i)+c(i)\,
{c^\dagger}^\alpha(i)\,c(i)\nonumber
\label{eq18}
\end{eqnarray}
$n_\mu(i)=c^\dagger(i)\,\sigma_\mu\,c(i)$ is the charge ($\mu=0$) and
spin ($\mu=1,2,3$) density operator for $c$-electrons. We are using the
following notation $\sigma_\mu=(1,\mathbf{\sigma})$,
$\sigma^\mu=(-1,\mathbf{\sigma})$, $\mathbf{\sigma}$ being the Pauli
matrices. The composite fields (\ref{eq17}) do not satisfy a canonical
algebra. For example, for the first two fields
\begin{eqnarray}
\left\{\xi(i),\xi^\dagger(j)\right\}&=&
\delta_{ij}\left[1-\frac12\sigma^\mu\,n_\mu(i)\right]\nonumber\\
\left\{\eta(i),\eta^\dagger(j)\right\}&=&
-\delta_{ij}\frac12\sigma^\mu\,n_\mu(i)\nonumber\\
\left\{\xi(i),\eta^\dagger(j)\right\}&=&\left\{\xi(i),\xi(j)\right\}
=\left\{\eta(i),\eta(j)\right\}
=0
\label{eq19}
\end{eqnarray}
This field satisfies the Heisenberg equation
\begin{equation}
i\frac{\partial}{\partial t}\psi(i)=J(i)=\left(
 \begin{array}{c}
  -\mu\,\xi(i)-4t\,c^\alpha(i)-4t\,\pi(i) \\ (-\mu+U)\eta(i)+4t\,\pi(i)
  \\ -\mu\,\pi(i)+4t\,\kappa(i)-4t\,\theta(i)+U\,\rho(i)
 \end{array}
\right)
\label{eq20}
\end{equation}
where
\begin{eqnarray}
\kappa(i)&=&\frac12\sigma^\mu\,c(i)\,c^{\dagger\alpha}(i)\,
\sigma_\mu\,c^\alpha(i)-\frac12\sigma^\mu\,c^\alpha(i)\,
c^\dagger(i)\,\sigma_\mu\,c^\alpha(i)\nonumber\\
\theta(i)&=&\frac12\sigma^\mu\,n_\mu(i)\,c^{\alpha^2}(i)
+c^\alpha(i)\,c^{\dagger\alpha}(i)\,c(i)
-c(i)\,c^{\dagger\alpha^2}(i)\,c(i)
+c(i)\,c^{\dagger\alpha}(i)\,c^\alpha(i)\nonumber\\
\rho(i)&=&\frac12\sigma^\mu\,n_\mu(i)\,\eta^\alpha(i)
+c(i)\,\xi^{\dagger\alpha}(i)\,c(i)
\label{eq21}
\end{eqnarray}

According to the method given above, the equation of motion (\ref{eq20})
is linearized as
\begin{equation}
i\frac{\partial}{\partial t}\psi(i)=\sum_j\varepsilon(i,j)\,\psi(j)
\label{eq22}
\end{equation}
where the energy matrix $\varepsilon(i,j)$ is the $3\times3$ matrix
given by
\begin{equation}
\varepsilon(i,j)=\left\langle\left\{J(i),\psi^\dagger(j)
\right\}\right\rangle_\mathrm{E.T.}
\left\langle\left\{\psi(i),\psi^\dagger(j)
\right\}\right\rangle_\mathrm{E.T.}^{-1}
\label{eq23}
\end{equation}
The subscript \textrm{E.T.} indicates that the anticommutators are
evaluated at equal time.

The physical properties can be described in terms of the thermal
retarded Green's function
\begin{equation}
S(i,j)=\left\langle\emph{R}\left[\psi(i)\,\psi^\dagger(i)\right]\right\rangle
=\frac{i\,a^2}{(2\pi)^3}\int_{\Omega_\mathrm{B}}d^2k\,d\omega\,
e^{i\,\mathbf{k}\cdot\left(\mathbf{R}_i-\mathbf{R}_j\right)-i\,
\omega\left(t_i-t_j\right)}\,S(\mathbf{k},\omega)
\label{eq24}
\end{equation}
where $\emph{R}$ is the usual retarded operator and the symbol
$\langle\cdots\rangle$ denotes the thermal average. By means of the
linearized Heisenberg equation (\ref{eq22}) the Fourier transform is
given by
\begin{equation}
S(\mathbf{k},\omega)
=\sum_{n=1}^3\frac{\sigma^n(\mathbf{k})}{\omega-E_n(\mathbf{k})+i\,\eta}
\label{eq25}
\end{equation}
where the energy spectra $E_i(\mathbf{k})$ are the characteristic values
of the matrix $\varepsilon(\mathbf{k})$, determined by the equation
\begin{equation}
\sum_{m=1}^{3}a_m(\mathbf{k})\,E_n^m(\mathbf{k})=0
\label{eq26}
\end{equation}

The characteristic coefficients $a_i(\mathbf{k})$ are defined by the
following relation
\begin{equation}
a_{n-k}(\mathbf{k})=(-)^k\emph{Tr}_k\left[\varepsilon(\mathbf{k})\right]
\hspace{1cm} 0 \le k \le 3
\label{eq27}
\end{equation}
where $\emph{Tr}_k$ is the trace of the $k^{th}$ order, defined as the
sum of the determinants of all $\left(\begin{array}{c}
  3\\k
\end{array}\right)$ matrices of order $k \times k$ which can be formed by intersecting
any $k$ rows of $\varepsilon$ with the same $k$ columns. We note that
$\emph{Tr}_3[\varepsilon]=\emph{Det}[\varepsilon]$ and the convention
$\emph{Tr}_0[\varepsilon]=1$ is used. The spectral functions are given
by
\begin{equation}
\sigma^n(\mathbf{k})=\frac1{b_n(\mathbf{k})}
\sum_{m=0}^2E_n^m(\mathbf{k})\,\lambda^m(\mathbf{k})
\label{eq28}
\end{equation}
where the $\lambda^n(\mathbf{k})$ are the $3 \times 3$ matrices:
\begin{equation}
\lambda^n(\mathbf{k})=\sum_{m=n+1}^3 a_m(\mathbf{k})\,
\varepsilon^{m-n-1}(\mathbf{k})\,I(\mathbf{k})
\hspace{1cm} 0 \le n \le 2
\label{eq29}
\end{equation}
and we put
\begin{equation}
b_n(\mathbf{k})=
\prod_{m=1,m \ne n}^3\left[E_n(\mathbf{k})-E_m(\mathbf{k})\right]
\label{eq30}
\end{equation}

By standard arguments, the correlation functions can be calculated from
the knowledge of the retarded Green's function. By means of (\ref{eq25})
we have
\begin{equation}
C(i,j)=\left\langle\psi(i)\,\psi^\dagger(j)\right\rangle
=\frac{a^2}{2(2\pi)^2}\sum_{n=1}^3 \int d^2k\,
e^{i\,\mathbf{k}\cdot\left(\mathbf{R}_i-\mathbf{R}_j\right)-i\,
E_n(\mathbf{k})\left(t_i-t_j\right)}\,
\sigma^n(\mathbf{k})\left[1+T_n(\mathbf{k})\right]
\label{eq31}
\end{equation}
where we put
\begin{equation}
T_n(\mathbf{k})
=\tanh\left(\frac{E_n(\mathbf{k})}{2k_\mathrm{B}\,T}\right)
\label{eq32}
\end{equation}

We see that the calculation of the Green's function requires the
knowledge of the normalization matrix
\begin{equation}
I(\mathbf{k})=\mathcal{F}\left\langle\left\{\psi(i),\psi^\dagger(j)
\right\}\right\rangle_{\mathrm{E.T.}}
\label{eq33}
\end{equation}
and of the $m$-matrix
\begin{equation}
m(\mathbf{k})=\mathcal{F}\left\langle\left\{J(i),\psi^\dagger(j)
\right\}\right\rangle_{\mathrm{E.T.}}
\label{eq34}
\end{equation}
where $\mathcal{F}$ indicates the Fourier transform.

These quantities are calculated in appendix and depends on a series of
parameters, that can be so listed:
\begin{enumerate}
 \item external parameters as the temperature $T$ and the electron
density $n=\left\langle c^\dagger(i)\,c(i) \right\rangle$; \item model
parameters as $U$ and $t$;
\item self-consistent parameters that can be calculated in terms of
elements of the Green's function, as $\mu$ and $\Delta$;
 \item self-consistent parameters expressed as expectation values of
composite fields out of the basis (\ref{eq17}), as $p$, $I_{33}^0$,
$I_{33}^\alpha$, $m_{33}^0$, $m_{33}^\alpha$.
\end{enumerate}
For the latter a procedure of self-consistency must be fixed. In the
Composite Operator Method (\emph{COM}) we take advantage of this freedom
and we fix the parameters in such a way that the Hilbert space has the
right properties to conserve the relations among matrix elements imposed
by symmetry laws. In a physics dominated by a high correlation among the
electrons, the first attention should be put to the requirement that the
approximation does not violate the symmetry required by the Pauli
principle. Let us consider the correlation matrix (\ref{eq31}); when we
take equal points the algebra leads to the following relations
\begin{eqnarray}
\left\langle\xi(i)\,\eta^\dagger(i)\right\rangle&=&0\nonumber\\
\left\langle\xi(i)\,\pi^\dagger(i)\right\rangle&=&
\left\langle c^\alpha(i)\,\xi^\dagger(i)\right\rangle\\
\left\langle\eta(i)\,\pi^\dagger(i)\right\rangle&=&
-\left\langle c^\alpha(i)\,c^\dagger(i)\right\rangle
\nonumber
\label{eq35}
\end{eqnarray}
among matrix elements of the Green's function. These relations
constitute a set of coupled self-consistent equations which will be
satisfied by an appropriate choice of the parameters.

The recovery of the Pauli principle does not exhaust all the degrees of
freedom and we have place to accommodate other symmetries. An intrinsic
symmetry of the Hubbard model is the pseudospin $SU(2)$ symmetry, which
is nothing also that the invariance under the particle-hole
transformation. The generators of this transformation are given by the
total pseudospin operators
\begin{eqnarray}
P^+&=&\sum_i(-)^i\,c_\uparrow^\dagger(i)\,c_\downarrow^\dagger(i)\nonumber\\
P^-&=&\sum_i(-)^i\,c_\downarrow(i)\,c_\uparrow(i)\\
P_z&=&\frac12\sum_i\left[n(i)-1\right]
\nonumber
\label{eq36}
\end{eqnarray}
These operators satisfy the $SU(2)$ algebra
\begin{equation}
[P^+,P^-]=2P_z \hspace{1cm} [P^\pm,P_z]=\mp P^\pm
\label{eq37}
\end{equation}
and the Heisenberg equations
\begin{eqnarray}
i\frac{\partial}{\partial t}P^\pm&=&\pm(2\mu-U)P^\pm\nonumber\\
i\frac{\partial}{\partial t}P_z&=&0
\label{eq38}
\end{eqnarray}

Let us consider the thermal retarded Green's function
\begin{equation}
P^{+-}(t-t^\prime)
=\left\langle\emph{R}\left[P^+(t)P^-(t^\prime)\right]\right\rangle
=\frac{i}{2\pi}\int_{-\infty}^{+\infty}d\omega\,
e^{-i\,\omega(t-t^\prime)}\,P^{+-}(\omega)
\label{eq39}
\end{equation}
By means of the equation of motion (\ref{eq38}) we obtain for the
correlation function
\begin{equation}
\frac1N\left\langle P^+(t)P^-(t^\prime)\right\rangle
=\frac{(n-1)e^{-i(2\mu-U)\left(t-t^\prime\right)}}
{1-e^{-\beta(2\mu-U)}}
\label{eq40}
\end{equation}
This is an exact result which relates the pseudospin correlation
function to the particle number $n$ and it is a manifestation of the
intrinsic symmetry.

Another important symmetry is given by the conservation of the current
density. By defining the charge $\rho(i)$ and current $\mathbf{j}(i) $
densities as
\begin{eqnarray}
\rho(i)&=&e\,c^\dagger(i)\,c(i)
\label{eq41}\\
\mathbf{j}(i)&=&-i\,t\,e\,a^2\,c^\dagger(i)
[\stackrel{\rightarrow}{\nabla}-\stackrel{\leftarrow}{\nabla}]c(i)
\label{eq42}
\end{eqnarray}
it is immediate to obtain by means of the Heisenberg equation
(\ref{eq14}) the conservation law
\begin{equation}
\nabla\cdot\mathbf{j}(i)+\frac{\partial}{\partial t}\rho(i)=0
\label{eq43}
\end{equation}

The symmetry content of the algebraic equation (\ref{eq43}) manifests at
level of observation as relations among matrix elements once a choice of
the physical space of states has been made. Indeed, by defining the
causal charge and current functions as
\begin{equation}
\chi_{ab}(i,j)=\left\langle\emph{T}\left[g_a(i)g_b(i)\right]\right\rangle
=\frac{i\,a^2}{(2\pi)^3}\int d^2k\,d\omega\,
e^{i\,\mathbf{k}\cdot(\mathbf{R}_i-\mathbf{R}_j)
-i\,\omega\left(t_i-t_j\right)}\,\chi_{ab}(\mathbf{k},\omega)
\label{eq44}
\end{equation}
where
\begin{equation}
g_a(i)=\left\{
\begin{array}{lcr}
\rho(i)&\mathrm{for}&a=0\\
j_x(i)&\mathrm{for}&a=x\\ j_y(i)&\mathrm{for}&a=y\\
\end{array}
\right.
\label{eq45}
\end{equation}
we can derive a series of Ward-Takahashi identities connecting
current-current, charge-current and charge-charge propagators. One of
those reads as follows
\begin{equation}
i\,a\,\omega\,\chi_{00}(\mathbf{k},\omega)
=\left[1-e^{-i\,k_x\,a}\right]\chi_{x0}(\mathbf{k},\omega)
+\left[1-e^{-i\,k_y\,a}\right]\chi_{y0}(\mathbf{k},\omega)
\label{eq46}
\end{equation}

In the static approximation of the Composite Operator Method the charge,
current, spin, pseudospin correlation function can be connected to
convolutions of single-particle propagators. This occurrence is related
to a linearized dynamics together with the choice of occupation
dependent electronic excitations as basic fields \cite{COM}. Once these
calculations have been performed, Eqs.~(\ref{eq35}), (\ref{eq40}) and
(\ref{eq46}) constitute a set of five coupled self-consistent equations
which can be satisfied by an appropriate choice of the five parameters
$p$, $I_{33}^0$, $I_{33}^\alpha$, $m_{33}^0$, $m_{33}^\alpha$.

We thus have a scheme of calculations in which it is possible to
recover, through a self-consistent calculation, a series of fundamental
symmetries by choosing a suitable Hilbert space.

Detailed calculations will be presented elsewhere.

\newpage

\appendix

\setcounter{equation}{0}
\renewcommand\theequation{\Alph{section}.\arabic{equation}}

\begin{center}
{\normalfont\Large\sf\bfseries Appendices}
\end{center}

\section{The normalization matrix}
From the definition (\ref{eq33}) and by means of the canonical algebra
for the $c$-electrons it is straightforward to see that for a
paramagnetic ground state the normalization matrix has the following
expression
\begin{equation}
I(\mathbf{k})=\left(
\begin{array}{ccc}
I_{11}(\mathbf{k})&0&I_{13}(\mathbf{k})\\
0&I_{22}(\mathbf{k})&I_{23}(\mathbf{k})\\
I_{13}(\mathbf{k})&I_{23}(\mathbf{k})&I_{33}(\mathbf{k})
\end{array}
\right)
\end{equation}
with
\begin{eqnarray}
I_{11}(\mathbf{k})&=&1-\frac n2\nonumber\\
I_{13}(\mathbf{k})&=&\Delta+\left(p-I_{22}\right)\alpha(\mathbf{k})\nonumber\\
I_{22}(\mathbf{k})&=&\frac n2\\
I_{23}(\mathbf{k})&=&-\Delta-p\,\alpha(\mathbf{k})\nonumber\\
I_{33}(\mathbf{k})&=&I_{33}^0+\alpha(\mathbf{k})
\,I_{33}^\alpha\nonumber
\end{eqnarray}

The quantities introduced in (A.2) are so defined
\begin{eqnarray}
n&=&\left\langle c^\dagger(i)\,c(i)\right\rangle
=2\left[1-\left\langle\xi(i)\,\xi^\dagger(i)\right\rangle
-\left\langle\eta(i)\,\eta^\dagger(i)\right\rangle\right]\nonumber\\
\Delta&=&\left\langle\xi^\alpha(i)\,\xi^\dagger(i)\right\rangle
-\left\langle\eta^\alpha(i)\,\eta^\dagger(i)\right\rangle\\
p&=&\frac14\left\langle n_\mu^\alpha(i)\,n_\mu(i)\right\rangle
-\left\langle\left[\xi_\uparrow(i)\,\eta_\downarrow(i)\right]^\alpha
\eta_\downarrow^\dagger(i)\,\xi_\uparrow^\dagger(i)\right\rangle
\nonumber
\end{eqnarray}
$n$ is the average number of electrons per site; $\Delta$ and $p$ are
static correlation function between nearest neighbor sites. In
particular, the parameter $p$ describes intersite charge, spin and pair
correlations. In the calculation of $I_{33}(\mathbf{k})$ only the
nearest neighbor contributions have been retained
\begin{equation}
\left\langle\left\{\pi(i),\pi^\dagger(j)\right\}\right\rangle
\cong\delta_{ij}\,I_{33}^0+\alpha_{ij}\,I_{33}^\alpha
\end{equation}

\setcounter{equation}{0}

\section{The $m$-matrix}
At first we note that time translational invariance requires the
$m$-matrix be hermitian
\begin{equation}
m(\mathbf{k})=\left(
\begin{array}{ccc}
m_{11}(\mathbf{k})&m_{12}(\mathbf{k})&m_{13}(\mathbf{k})\\
m_{12}(\mathbf{k})&m_{22}(\mathbf{k})&m_{23}(\mathbf{k})\\
m_{13}(\mathbf{k})&m_{23}(\mathbf{k})&m_{33}(\mathbf{k})
\end{array}
\right)
\end{equation}

From the definition (\ref{eq34}) and by making use of the expressions
(\ref{eq20}--\ref{eq21}) for the source current it is possible to
calculate
\begin{eqnarray}
m_{11}(\mathbf{k})&=&-\left[\mu+4t\,
\alpha(\mathbf{k})\right]I_{11}-4t\,I_{13}\nonumber\\
m_{12}(\mathbf{k})&=&-4t\,
\alpha(\mathbf{k})\,I_{22}-4t\,I_{23}\nonumber\\
m_{13}(\mathbf{k})&=&-\left[\mu+4t\,\alpha(\mathbf{k})\right]I_{13}
-4t\left[\alpha(\mathbf{k})\,I_{23}+I_{33}\right]\\
m_{22}(\mathbf{k})&=&-(\mu-U)I_{22}+4t\,I_{23}\nonumber\\
m_{23}(\mathbf{k})&=&-(\mu-U)I_{23}+4t\,I_{33}\nonumber\\
m_{23}(\mathbf{k})&=&m_{33}^0+
\alpha(\mathbf{k})\,m_{33}^\alpha\nonumber
\end{eqnarray}
In the calculation of $m_{33}(\mathbf{k})$ only the nearest neighbor
contributions have been retained
\begin{equation}
\left\langle\left\{J_3(i),\pi^\dagger(j)\right\}\right\rangle
\cong\delta_{ij}\,m_{33}^0+\alpha_{ij}\,m_{33}^\alpha
\end{equation}

\end{document}